\documentclass[prc,twocolumn,showpacs,showkeys,nofootinbib,superscriptaddress]{revtex4}
\usepackage{graphicx}
\usepackage{dcolumn}
\usepackage{bm}

\begin{document}

\topmargin -0.50in

\title{Exact relativistic tritium $\beta$-decay endpoint spectrum
 in a hadron model}

\author{Fedor \v Simkovic}
\affiliation{Department of Nuclear Physics, 
Comenius University, Mlynsk\'a dolina F1, SK--842 15
Bratislava, Slovakia}
\affiliation{Institut f\"{u}r Theoretische Physik der Universit\"{a}t
T\"{u}bingen, D-72076 T\"{u}bingen, Germany}
\author{Rastislav Dvornick\'y}
\affiliation{Department of Nuclear Physics, 
Comenius University, Mlynsk\'a dolina F1, SK--842 15
Bratislava, Slovakia}
\author{Amand Faessler}
\affiliation{Institut f\"{u}r Theoretische Physik der Universit\"{a}t
T\"{u}bingen, D-72076 T\"{u}bingen, Germany}

\begin{abstract}
We present the relativistic calculation of the 
$\beta$-decay of tritium in a hadron model. The 
elementary particle treatment (EPT) of the transition
$^{3}H \rightarrow ^{3}He + e^- + \overline\nu_e$
is performed in analogy with the description of the 
$\beta$-decay of neutron. The effects of higher order 
terms of hadron current and nuclear recoil are 
taken into account in this formalism. The relativistic 
Kurie function is derived and presented 
in a simple form suitable for the determination of 
neutrino masses from the shape of the endpoint spectrum. 
A connection with the commonly used Kurie function 
is established.
\end{abstract}



\date{\today}

\maketitle

\section{Introduction}

Neutrinos are one of the most intriguing and fascinating 
fundamental particles, which make up the Universe. However, they 
are also one of the least understood particles. Studies of 
neutrinos have played a crucial role in the understanding 
of elementary particle laws and their interactions. 

Three types of light neutrinos are known. The recent observation 
of neutrino oscillations \cite{SK,SNO,Kamland,K2K,Minos} 
has now beyond doubt established the non-zero masses of neutrinos, 
the flavor change and neutrino mixing. It has opened a 
new excited era in neutrino physics and represents a big step 
forward in our knowledge of neutrino properties and serves as solution 
of many problems in cosmology, elementary particle physics, 
and astrophysics.

While neutrino oscillation experiments are sensitive only to 
differences of squared neutrino masses, the neutrino mass 
measurements with tritium ($Q_{\beta}(^{3}H) = 18.6~keV$)
and rhenium ($Q_{\beta}(^{187}Re) = 2.47~keV$) $\beta$-decays yield 
direct information on the absolute neutrino mass scale.
The idea underlying the measurement of  neutrino mass is 
actually fairly obvious. A long time ago, it was already
pointed out by E. Fermi \cite{fermi} 
that the shape of the electron spectrum in nuclear $\beta$-decay, 
near the kinematical end point, is sensitive to the neutrino mass.

Attempts to evaluate the rest mass of the neutrino experimentally 
were already being undertaken long ago.  In 1940 one of the first 
kinematical measurements of neutrino mass was performed by Hanna 
and Pontecorvo \cite{hanna} with a proportional chamber filled with tritium.
A limit of $\sim ~1~keV$ on the neutrino mass was obtained, which
was determined by the resolution of the detector.
The Mainz \cite{mainz} and  Troitsk \cite{troitsk} tritium $\beta$-decay 
experiments using the magnetic adiabatic collimation technique, 
place the present upper limit on the mass of the electron 
neutrino of $2.3~ eV$ and $2.2~ eV$, respectively. 
The best published calorimetric limit to the electron neutrino mass 
obtained from the $\beta$-spectrum of $^{187}Re$ is $15~eV$ \cite{mibeta}.
We note that the bounds on neutrino mass imposed by the shape of the spectrum 
are independent of whether neutrino is a Majorana or a Dirac particle.

A next-generation tritium $\beta$-decay experiment is the
KArlsruhe TRItium Neutrino experiment (KATRIN) \cite{katrin,drexlin,wein}, 
which is presently in construction phase (It is planned to take data 
starting 2010).  
This experiment is projected for measurement of the neutrino mass 
with a sensitivity of 200 meV, which will have important implications 
for the theory of neutrino masses. If the result will be positive, 
it will imply a degenerate spectrum of neutrino masses. On the other hand, 
a negative result will be a very useful constraint. There is also 
a chance that the planned MARE experiment \cite{mare} based on 
arrays of rhenium low temperature microcalorimeters will be able to achieve 
sensitivity  lower than 0.2 eV in future. The MARE approach would 
have totally different systematics with respect to the KATRIN.  

In view of an enormous experimental progress in the field 
there is a request for a highly accurate theoretical description 
of the electron energy spectrum in the determination of the neutrino masses 
from the shape of the endpoint spectrum. The subject of interest has been
molecular effects in tritium beta decay \cite{Doss}, 
radiative corrections \cite{meissner}, 
Lorentz invariance violations \cite{lorentz}, interactions beyond the
standard model \cite{goldman}, relativistic form for 
the $\beta$-decay endpoint spectrum  \cite{repko,masood} etc.  

The aim of this paper is to derive the relativistic form for 
the $\beta$-endpoint spectrum in a hadron model.
We shall take advantage of the fact that the nuclei $^{3}H$ and $^{3}He$ 
are, respectively,  the nuclear analogs of the neutron and the proton, 
i.e., they form an isospin SU(2) doublet. A correspondence to the 
commonly used  formulae will be established. We note that the
considered approach is known also as Elementary Particle Treatment
(EPT) of weak processes, which was developed by Kim and Primakoff 
\cite{kim}.

\section{The nuclear physics description of tritium $\beta$-decay}

By neglecting neutrino mixing for simplicity and taking into account
only left-handed weak interaction, the electron energy
spectrum for tritium $\beta$-decay is 
\begin{eqnarray}
N(E_e) &=& \frac{d \Gamma}{d E_e} \nonumber\\ 
  &=& 
\frac{G^2_F V^2_{ud}}{2 \pi^3} |M.E.|^2
F(Z,E_e) p_e E_e  \nonumber\\
&&\times 
\left( E_0 - E_e\right)
\sqrt{(E_0 - E_e)^2 - m^2_\nu},
\label{eq:1}
\end{eqnarray}
where $G_F$ is the Fermi constant and $V_{ud}$ is the element of the 
Cabbibo-Kobayashi-Maskawa (CKM) matrix. $p_e$, $E_e$ and $E_0$ 
are the momentum, energy, and maximal endpoint energy (in the case of
zero neutrino mass) of the electron, respectively.
$F(Z,E)$ denotes the relativistic Coulomb factor.

The transition is superallowed, a mix of Fermi and Gamow-Teller
transitions.  The absolute square of the nuclear matrix element
is given by
\begin{equation}
|M.E.|^2 =  f^2_V |M_F|^2 + f^2_A |M_{GT}|^2, 
\end{equation}
where the Fermi  and Gamow-Teller  matrix elements
take the form
\begin{eqnarray}
M_F &=& <^{3}He| \sum^3_{k=1} \tau^+_k |^{3}H>, \\
{\vec{M}}_{GT} &=& <^{3}He| \sum^3_{k=1} \tau^+_k \vec\sigma_k |^{3}H>.
\label{eq:2}
\end{eqnarray}
$f_V$ and $f_A$ are the vector and the axial-vector coupling 
constants of the nucleon, respectively. We note that the derivation of the 
differential decay rate in (\ref{eq:1}) involves non-relativistic
approximations and that only the $s_{1/2}$ states
of outgoing leptons are taken into account. 

The Fermi matrix element can be evaluated by assuming the exact isospin symmetry 
as well as the fact that $^{3}H$ and $^{3}He$ form an isospin doublet $(T=1/2)$ 
(the projection $T_z = 1/2$
is assigned to the $^{3}He$ and $T_z = -1/2$ to the $^{3}H$) with the result
$M_F = 1$.  

The absolute square of the Gamow-Teller matrix element can be deduced 
from the Ikeda sum rule by taking into account 
that the Gamow-Teller operator has no radial dependence and thus can  
not scatter into higher shells. In $^{3}He$ the 1s neutron level
is already occupied by two neutrons and therefore in the transition $p$ to $n$ 
the neutron would need to be scattered into a higher orbit (e. g., 2s) 
in the continuum, which is forbidden for the Gamow-Teller operator. 
Thus only  $^{3}H \rightarrow ^{3}He$  but not 
$^{3}H \rightarrow 3n$ can contribute  to  the Ikeda sum rule.
In addition, there are no excited states of $^{3}He$. 
As a consequence $|M_{GT}|^2 = 3$. This result is in
a good agreement with the recommended  value $|M_{GT}| = \sqrt{3} (0.962\pm 0.002)$
obtained in nuclear structure calculation \cite{brown}.

The conserved vector current (CVC) hypothesis proposed by Feynman and Gell-Mann 
suggests that the vector coupling constant $f_V$ is not renormalized in the nuclear 
medium, i.e., $f_V = 1.0$. The accurately measured $\beta$-decay lifetime of tritium 
($T_{1/2} (^{3}H) = 12.32 \pm 0.03~years$) \cite{budick,hft} is used to adjust 
the value of  axial-vector coupling constant $f_A$ via the calculation
of the theoretical half-life
\begin{equation}
\left(T_{1/2}\right)^{-1/2} = \frac{\Gamma}{\ln{2}} = \int_{m_e}^{E_0 - m_\nu} N(E_e) dE_e. 
\end{equation}
In the computation of the integral over the electron energy $E_e$ we adopted 
the relativistic Coulombic factor $F(Z,E)$ \cite{doi}, which take into account 
the finite size of the nucleus. For $|M_{GT}|^2 = 3$ we found $|f_A| = 1.247$. 
The very good agreement between this result and the bare nucleon value 
$|f_A /f_V | = 1.2695 \pm 0. 0029$ \cite{PDG} suggests that the axial-vector 
coupling constant is only weakly quenched in the tritium.

The dependence of spectrum shape on the mass of neutrino $m_\nu$ in (\ref{eq:1}) 
follows from the phase volume factors only. The traditional way to look at 
the $\beta$-spectrum data is to make a Kurie plot, where 
\begin{eqnarray}
K (E_e) &\equiv & \sqrt{\frac{N(E_e)}{F(Z,E_e) p_e E_e}} \nonumber\\ 
        &=& \frac{G_F V_{ud}}{\sqrt{2 \pi^3}}  |M.E|\nonumber\\
&&\times \left( E_0 - E_e\right) 
\sqrt[4]{1 - \left( \frac{m_\nu}{(E_0 - E_e)}\right)^2}.
\label{eq:3}
\end{eqnarray} 
For zero mass neutrino, if $K(E_e)$ is plotted against $E_e$,
the result is a straight line that crosses the $E_e$ axis at $E_e = E_0$.
For $m_\nu \ne 0$ the endpoint shifts to $E^{max} = E_0 - m_\nu$ and 
the rate near the endpoint is depressed, namely  the Kurie 
plot has a kink at the endpoint. This distortion will be washed out
at the experiment unless the energy resolution is comparable to $m_\nu$.

There are open questions related to the  presented conventional approach 
for kinematical study of the $\beta$-decay endpoint of $^{3}H$. 
In particular, it is not known what the consequences of the 
considered non-relativistic approximations are. Further, the effect of the 
nuclear recoil is not taken into account. It is also worth mentioning
that the relativistic expression for the maximal electron energy 
\begin{equation}
E^{max}_e = \frac{1}{2 M_f}\left[ M^2_i + m^2_e - (M_f + m_\nu)^2 \right],
\label{eq:4}
\end{equation} 
gives a value about $3.4~eV$ lower than the considered approximation 
$E^{max}_e \simeq M_i - M_f - m_\nu$ \cite{masood} ($M_i$, $M_f$ and $m_e$ are
masses of the tritium atom, $^{3}He^+$ and the electron, 
respectively). In view of the planned sensitivity  of $\sim 0.2~eV$ 
of the KATRIN experiment, there is a request for a consistent relativistic 
description of the $\beta$-decay of tritium \cite{masood}.  
 
\section{Relativistic $\beta$-decay kinematics in hadron model}

We shall study the $\beta$-decay of tritium,
\begin{equation}
^3H~ \rightarrow ~^3He~ + ~e^-~ + {\overline\nu}_e,
\label{eq:5}
\end{equation}
in an analogy with the $\beta$-decay of a free neutron,
\begin{equation}
n~ \rightarrow ~p~ + ~e^-~ + {\overline\nu}_e,
\label{eq:6}
\end{equation}
as the spin-isospin characteristics of $^3H$ ($^3He$) nucleus 
and neutron (proton) are the same.  
The kinematics of the two processes above differ mostly due
to different Q-values  and the Coulomb corrections.

The invariant $\beta$-decay amplitude is given by
\begin{eqnarray}
{M} &=& \frac{G_F V_{ud}}{\sqrt{2}}~
\overline{u}(P_e)\gamma_\alpha (1 - \gamma_5) v(P_\nu)
\nonumber\\
&& \times \overline{u} (P_f) \left[ G_V(q^2) \gamma^\alpha 
+i \frac{ G_M(q^2)}{2 M_i} 
\sigma^{\alpha\beta} q_\beta \right.
\nonumber\\
&&\left.
~~~~~~~~~ - G_A(q^2)\gamma^\alpha \gamma_5
- G_P(q^2) q^\alpha \gamma_5 \right] u(P_i).
\nonumber\\
\label{eq:7}
\end{eqnarray} 
Here, $q_\alpha = (P_f-P_i)_\alpha = (P_e + P_\nu)_\alpha$ 
is the momentum transferred to the hadron vertex. 
$P_i = (M_i,0)$, $P_f = (M_f,{\mathbf{p_f}})$, $P_e = (m_e,{\mathbf{p_e}})$ 
and $P_\nu = (m_\nu,{\mathbf{p_\nu}})$ are four momenta 
of the  $^3H$, $^3He$, electron and antineutrino in the
laboratory frame, respectively.

The form factors $G_V(q^2)$, $G_M(q^2)$, $G_A(q^2)$, $G_P(q^2)$ 
are real functions of the squared momentum $q^2$. They are 
parameterized as follows:
\begin{eqnarray}
G_{V}(q^2) &=&
\frac{g_V}{\left(1-\frac{q^{2}}{M^2_V}\right)^2},~~~~
G_{M}(q^2) =
\frac{g_M}{\left(1-\frac{q^{2}}{M^2_V}\right)^2}, \nonumber\\
G_{A}(q^2) &=&
\frac{g_A}{\left(1-\frac{q^{2}}{M^2_A}\right)^2} . 
\end{eqnarray}
The two form-factor cut-offs $M_V$ and $M_A$ are in general different 
and their values are expected to be of the order of $1~GeV$ like it
is in the case of nucleon form-factors. As it will be discussed later
the $q^2$-dependence of these form-factors is not crucial for tritium 
$\beta$-decay. 

The conserved vector current hypothesis (CVC) implies $g_V = 1.0$.  
$g_M = -6.106$ is calculated from the values of magnetic moments 
of $^{3}H$ and $^{3}He$ using the CVC hypothesis as well \cite{stone}. 
The axial coupling constant $g_A$ can be determined from 
the measured half-life of $^{3}H$. The induced pseudoscalar 
coupling is given by the partially conserved
axial-vector current hypothesis (PCAC)
\begin{equation}
g_P (q^{2}) = {2 M_i g_A(q^{2})}/({m^2_\pi - {q}^{2}}).
\end{equation}
$m_\pi$ is the mass of pion.

For the spin-summed, Lorentz-invariant squared amplitude we get
\begin{eqnarray}
\frac{1}{2} \sum_{spins} |{M}|^2 =   
16 (G_F V_{ud})^2~~~~~~~~~~~~\nonumber\\
\times \left[ G^2_V {\cal P}_{VV} + G_A G_V {\cal P}_{AV} 
+ G^2_A {\cal P}_{AA} + 
~~~~~\right. \nonumber\\
+ G_A G_P {\cal P}_{AP} + G^2_P {\cal P}_{PP}~~~~~~~~~~~~~ 
\nonumber\\
\left.
+ G_V G_M \frac{{\cal P}_{VM}}{2M_i}  
+  G_A G_M \frac{{\cal P}_{AM}}{2 M_{i}} +  G_M^2 \frac{{\cal P}_{MM}}{4 M^2_{i}}
\right]\nonumber\\
\label{eq:8}
\end{eqnarray}
with

\begin{equation}
{\cal P}_{VV} = P_{ef} P_{\nu i} + P_{e i} P_{\nu f} - M_i M_f P_{e\nu}, 
\end{equation}
\begin{equation}
{\cal P}_{AA} = P_{ef} P_{\nu i} + P_{e i} P_{\nu f} + M_i M_f P_{e\nu}, 
\end{equation}
\begin{equation}
{\cal P}_{AV} = 2 \left( P_{ef} P_{\nu i} - P_{e i} P_{\nu f} \right), 
\end{equation}
\begin{equation}
{\cal P}_{AP} =
M_f ( m^2_e P_{\nu i} + m^2_\nu P_{ei} ) -
M_i ( m^2_e P_{\nu f} + m^2_\nu P_{ef} ),
\end{equation}
\begin{equation}
{\cal P}_{PP}~ =~\frac{1}{2}(P_{if} - M_i M_f) 
\left( P_{e\nu} (m^2_e + m^2_\nu) + 2 m^2_\nu m^2_e \right), 
\end{equation}
\begin{eqnarray}
{\cal P}_{VM} =~~~~~~~~~~~~~~~~~~~  
~~~~~~~~~~~~~~~~~
\nonumber\\
~ M_i  \left[  P_{e\nu} (P_{if}- M^2_f  ) 
+ P_{ef}
(P_{\nu i} -2 P_{\nu f} ) + P_{e i} P_{\nu f} 
\right]\nonumber\\
+M_f \left[  P_{e\nu} (P_{if}- M^2_i) 
+ P_{ei}
(P_{\nu f} -2 P_{\nu i} ) + P_{e f} P_{\nu i} 
\right],
\end{eqnarray}
\begin{equation}
{\cal P}_{AM} = 2 (M_i + M_f) ( P_{ef} P_{\nu i} - P_{e i} P_{\nu f}),
\end{equation}
\begin{eqnarray}
{\cal P}_{MM}~ =
~~~~~~~~~ ~~~~~~~~~~~~~~~~~~~~~~~~~\nonumber\\
- \frac{1}{2} P_{if} 
\left( P_{e\nu} ( m^2_e + m^2_\nu ) + 2 m^2_e m^2_\nu \right) 
- M_i M_f m^2_e m^2_\nu
\nonumber\\
+ 2 P_{e i} P_{e f} (P_{e \nu} + m^2_\nu ) 
+ 2 P_{\nu i} P_{\nu f} (P_{e \nu} + m^2_e ) 
~~~~~~~ 
\nonumber\\
-\frac{1}{2} M_i M_f  P_{e \nu} \left( 3 m^2_e + 3 m^2_\nu  + 4 P_{e\nu} \right).
~~~~~~~~~~~~~
\label{eq:9}
\end{eqnarray}
Here, $P_{kl} \equiv \left(P_k\cdot P_l\right)$  with $k,l = i,~f,~e$ and $\nu$ denotes the
scalar product of two four-momenta. 

By neglecting the contribution from higher order currents (terms proportional to
$G_{M,P}$) we find
\begin{eqnarray}
\frac{1}{2} \sum_{spins} |{M}|^2 =   
16 (G_F V_{ud})^2~~~~~~~~~~~~\nonumber\\
\times \left[ 
(G_V + G_A)^2 (P_e \cdot P_f) (P_\nu \cdot P_i) 
~~~~~~\right.\nonumber\\
+ (G_V - G_A)^2 (P_e \cdot P_i) (P_\nu \cdot P_f) 
~~~~~~ \nonumber\\
\left.
(-G^2_V + G^2_A) M_i M_f (P_e \cdot P_\nu)
\right].~~~~~~
\label{eq:10}
\end{eqnarray}

The advantage of the presented formalism is that the squared Lorentz invariant 
amplitude is calculated exactly unlike in Ref. \cite{masood}, where an assumption 
about its dominant constituent was considered. We note that for $G_V = G_A =1$ the 
squared amplitude is proportional to $(P_e \cdot P_f) (P_\nu \cdot P_i)$, i.e., 
the structure is similar as, e.g., in the case of the muon decay.

For the  tritium $\beta$-decay at rest 
the differential decay rate is 
\begin{eqnarray}
d\Gamma &=& 
\frac{1}{2 M_i} F(Z,E_e) 
\left( \frac{1}{2} \sum_{spins} |{M}|^2 \right)
\nonumber\\
&&\times\frac{(2\pi)^4}{(2\pi)^9} 
\delta^{(4)}(P_i - P_f - P_e - P_{\nu})
\frac{d^3 p_e}{2 E_e} \frac{d^3 p_\nu}{2 E_\nu} 
\frac{d^3 p_f}{2 E_f}.\nonumber\\
\label{eq:11}
\end{eqnarray}
The factor $1/2$ in front of the squared amplitude stands 
for the average over the spin of 
the initial state. 

The subject of interest is the energy distribution of the electron.
Hence, the integration over antineutrino and final nucleus momenta
have to be performed in (\ref{eq:11}). It requires calculation of the following 
integrals:  
\begin{eqnarray}
{\cal K} &=& \int 
\frac{d^3 p_f}{E_f}
\frac{d^3 p_\nu}{E_\nu} 
~\delta^{(4)}(Q - P_f - P_{\nu}), \\
({\cal L}_{\nu,f})^{\rho}
&=& \int 
\frac{d^3 p_f}{E_f}
\frac{d^3 p_\nu}{E_\nu} 
~\delta^{(4)}(Q - P_f - P_{\nu}) (P_{\nu, f})^\rho, \\
({\cal N}_{kl})^{\rho\sigma} &=&  \int
\frac{d^3 p_f}{E_f}
\frac{d^3 p_\nu}{E_\nu} 
~\delta^{(4)}(Q - P_f - P_{\nu}) (P_k)^\rho (P_l)^\sigma 
\nonumber\\
\label{eq:12}
\end{eqnarray}
with $Q=P_i -P_e$ and $k,l = \nu, f$. The details
of integrations with results are given in the Appendix.

The differential decay rate is found to be of the form
\begin{eqnarray}
\frac{d\Gamma}{d E_e} &=& \frac{1}{2 \pi^3}
(G_F V_{ud})^2 F(Z,E_e) p_e 
\nonumber\\
&& \times \frac{M^2_i}{(m_{12})^2}
\sqrt{y \left(  y + 2 m_\nu \frac{M_f}{M_i} \right)}
\nonumber\\
&&\times \left[
g^2_V {\cal R}_{VV} + g_A g_V {\cal R}_{AV} 
+ g^2_A {\cal R}_{AA} + 
\right. \nonumber\\
&&~~~~~~~+ g_A g_P {\cal R}_{AP} + g^2_P {\cal R}_{PP} 
\nonumber\\
&&~~\left.
+ g_V g_M {\cal R}_{VM}
+  g_A g_M {\cal R}_{AM}
+  g_M^2 {\cal R}_{MM}
\right],\nonumber\\
\label{eq:13}
\end{eqnarray}
where $(m_{12})^2 = M_i^2 + m_e^2 - 2 M_i E_e$ and $y = E^{max}_e - E_e$.
In the calculation we neglected $q^2$ dependence of the form-factors as 
for the $\beta$-decay of $^{3}H$ the value of $q^2$ is rather small.
Their consideration would lead only to small correction factors, which
are not sensitive to neutrino mass. We find not usefull to present here 
the explicit form of all ${\cal R}_I$ ($I=VV,~VA,~AA,~AP,~PP,~VM,~AM,~MM$) 
factors. Instead of that we conclude about their structure and importance. 

Our analysis showed that each term of  ${\cal R}_I$ is proportional
to $(y + m_\nu ({M_f + m_\nu})/{M_i})$ or  $(y + m_\nu {M_f}/{M_i})$. 
So, a common $(y + m_\nu {M_f}/{M_i})$ can be put in front of the bracket
in (\ref{eq:13})
by neglecting a small term $m_\nu /M_i$. The importance of 
different ${\cal R}_I$ contributions can be studied 
in the limit $M_i=M_f$, $E_e=m_e$ and by making Taylor 
expansion in in $m_\nu$, $m_e$ ($m_\nu \ll m_e \ll M_i$).
The leading terms of  different ${\cal R}_I$ 
(without the common factor) are as follows:
\begin{eqnarray}
VV:~m_e M_i,~~~AA:~3 m_e M_i,~~~AV:~2 m^2_e,~~~\nonumber\\
VM:~\frac{1}{2}\frac{m^3_e}{M_i},~~~
MM:~\frac{3}{16}\frac{m^5_e}{M^3_i},~~~
AM:~2 m^2_e,~~\nonumber\\
AP:~2 m_e M_i \frac{m^2_e}{m^2_\pi},~~
PP:~\frac{1}{2}m_e M_i \frac{m^4_e}{M^2_i m_\pi^2}.~~~~~~~
\nonumber\\
\label{eq:14}
\end{eqnarray}
From their comparison we conclude that the contributions coming from 
higher order terms of hadron current to the decay rate of $^{3}H$
can be neglected.

\begin{figure}[tb]
\includegraphics[width=.42\textwidth]{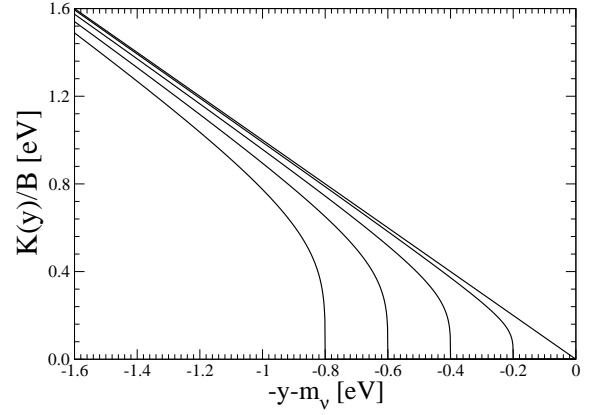}
\caption{Endpoints of the relativistic 
Kurie plot [see Eqs. (\ref{eq:17}) and (\ref{eq:18})]
of the tritium beta decay for various values of the neutrino mass:
$m_\nu = 0,~0.2,~0.4,~0.6,$ and $0.8~eV$.
}
\label{fig:1}
\end{figure}

Then we have 
\begin{eqnarray}
\frac{d\Gamma}{d E_e} &=& \frac{1}{2 \pi^3}
(G_F V_{ud} )^2 F(Z,E_e) p_e 
\nonumber\\
&& \times \frac{M^2_i}{(m_{12})^2}
\sqrt{y \left(  y + 2 m_\nu \frac{M_f}{M_i} \right)}
\nonumber\\
&&\times 
\left[
(g_V + g_A)^2 
y \left( y + m_\nu \frac{M_f}{M_i} \right) 
\frac{ M_i^2 (E^2_e-m^2_e) }{ 3 (m_{12})^4} 
\right. 
\nonumber\\
&&~~(g_V + g_A)^2 (y + m_\nu \frac{M_f + m_\nu}{M_i})
\frac{(M_i E_e - m_e^2)}{m_{12}^2}  
\nonumber\\
&&~~~~~~~~~~~~ \times (y + M_f \frac{M_f + m_\nu}{M_i}) 
\frac{(M_i^2 - M_i E_e)}{m_{12}^2} 
\nonumber\\
&& ~-
(g_V^2- g^2_A)  M_f
\left( y + m_\nu \frac{(M_f + M_\nu)}{M_i} \right)
\nonumber\\
&&~~~~~~~~~~~~~~~~~\times \frac{(M_i E_e - m^2_e)}{(m_{12})^2} 
\nonumber\\
&&~ \left. + 
(g_V - g_A)^2 E_e \left( y + m_\nu \frac{M_f}{M_i}\right) 
\right] .
\label{eq:15}
\end{eqnarray}
The first term in the brackets in (\ref{eq:15}), which is quadratic 
in y, plays a subleading role. By keeping only the dominant contributions
and by introducing a mass scale parameter $M$ instead of the $M_i$ and $M_f$,
we get
\begin{eqnarray}
\frac{d\Gamma}{d E_e} &\simeq & \frac{1}{2 \pi^3}
(G_F V_{ud})^2 F(Z,E_e) p_e E_e (g^2_V + 3 g^2_A) 
\nonumber\\
&& \times 
\sqrt{y \left(  y + 2 m_\nu  \right)} \left( y + m_\nu \right).
\label{eq:16}
\end{eqnarray} 

For the relativistic form of the Kurie function we can 
write
\begin{eqnarray}
K (y) = B \left(
\sqrt{y \left(  y + 2 m_\nu  \right)} \left( y + m_\nu \right)
\right)^{1/2}
\label{eq:17}
\end{eqnarray}
with 
\begin{equation}
B =  \frac{G_F V_{ud}}{\sqrt{2 \pi^3}} \sqrt{g^2_V + 3 g^2_A}. 
\label{eq:18}
\end{equation}
The unknown coupling constant $g_A$ of the hadron current is fixed 
to the half-life of ${^{3}H}$ \cite{budick,hft}  with result 
$g_A = 1.247$. This value coincides well with that of the 
axial-vector coupling of the nucleon (see previous section). 
We have $B = 3.43\times 10^{-6}~GeV^{-2}$.

By comparing the Kurie function in (\ref{eq:17}) and (\ref{eq:18}) 
with the commonly used one (\ref{eq:3}) we find that they are equal
if $y$ is replaced with $(E_0 - E_e - m_\nu)$ and $|M_{GT}|^2 = 3$
is assumed. This confirms what was generally expected, namely that 
the relativistic effects are small corrections to the results 
known in the traditional method due to a small $Q$-value of the 
$\beta$-decay of tritium. However, it was not clear yet whether the 
recoil of the nucleus, which value is $3.4$ eV for maximal electron
energy, affects the endpoint spectra, if sub eV mass of neutrino
is measured. Within the considered EPT of $\beta$-decay of tritium 
we find that there is no significant modification of the shape of
the electron spectra close to the endpoint due to 
the nuclear recoil.

In Fig. \ref{fig:1} we show a relativistic Kurie plot 
for the $\beta$-decay of $^{3}H$ versus $y = E^{max}-E_e$ 
near the endpoint. Special attention is given to 
the effect of a small neutrino mass ($m_\nu = 0.2,~0.4,~0.6$ and $0.8~eV$).
We see that the Kurie plot is linear near the endpoint 
for zero neutrino mass ($m_\nu=0$). However, the linearity 
of the Kurie plot is lost 
if the neutrino has a non-zero mass. Deviation from a straight line depends 
on  the magnitude of neutrino mass $m_\nu$. Though, there is no 
difference with the previously known dependences, it is worth to stress 
that in this case the relativistic form of the $\beta$-decay Kurie plot
is used, which also takes the nuclear recoil ($\sim 3.4$ eV) into 
account.

\section{Conclusion}

The neutrino absolute mass scale, which is very important for particle 
physics as well as for cosmology and astrophysics, cannot be resolved 
by oscillation experiments. A way of the direct determination of 
the neutrino mass scale in laboratory experiment is the investigation 
of the kinematics of tritium $\beta$-decay. 

The KATRIN experiment \cite{katrin,drexlin,wein}, which is under construction, 
will be able to reach  a sensitivity of  neutrino mass in the sub-eV range.
In connection with that there is a request for a highly accurate 
theoretical description of the electron energy spectrum.

In this paper we derived the relativistic form for the $\beta$-decay
endpoint spectrum in the elementary particle treatment of  weak
interaction. The considered formalism follows from the analogy 
between $^{3}H$ ($^{3}He$) and the neutron (proton) having the same
spin-isospin properties. It allowed us unlike 
in Ref. \cite{masood} to determine the squared $\beta$-decay amplitude 
more accurately. In addition, we found that the higher order terms of the 
hadron current can be neglected without affecting the dependence 
of the Kurie plot on the electron energy and the neutrino mass. 
By comparing the relativistic and previously used Kurie functions 
a good agreement between them was established. 

We acknowledge the support of  the EU ILIAS project under the contract
RII3-CT-2004-506222, the Deutsche Forschungsgemeinschaft (436 SLK 17/298)
and of the VEGA Grant agency of the Slovak Republic under the contract 
No.~1/0249/03. 

\appendix
\section{}
Here we outline the calculation of integrals over neutrino and
final nuclear momenta.  

\vspace{0.2cm}

{\bf Integration of ${\cal K}$}:\\
The integration is performed by choosing $Q = (Q_0,\mathbf{0})$,
i.e., the rest frame connected with the center of mass of antineutrino and
final nucleus. We have
\begin{eqnarray}
{\cal K} &=& \int \int
\frac{d^3 p_f}{E_f}
\frac{d^3 p_\nu}{E_\nu} 
~\delta^{(4)}(Q - P_f - P_{\nu})\nonumber\\
  &=& \int \frac{1}{E_\nu} 
~\delta(Q_0 - E_f - E_{\nu})
~{p_f  d E_f d \Omega_f}
\nonumber\\
\end{eqnarray}
with $E_f = (m_\nu^2 - M^2_f + E_f^2)^{1/2}$. By using 
$\delta (f(x)) = \delta(x-x_0)/{|f'(x_0)|}$ we find
\begin{eqnarray}
{\cal K} = 2\pi \frac{\sqrt{
[Q^2_0 - (M_f + m_\nu)^2]
[Q^2_0 - (M_f - m_\nu)^2]
}
}{Q^2_0}. 
\nonumber\\
\end{eqnarray}
We replace $Q_0^2$  with $Q^2$ and write ${\cal K}$
in the Lorentz invariant form 
\begin{eqnarray}
{\cal K} &=& 2\pi \frac{\sqrt{[Q^2 - (M_f + m_\nu)^2] [Q^2 - (M_f - m_\nu)^2]}
}{Q^2} \nonumber\\
&=& 4\pi M_i 
\frac{\sqrt{y \left(  y + 2 m_\nu \frac{M_f}{M_i} \right)}}
{(m_{12})^2}.
\end{eqnarray}

\vspace{0.2cm}

{\bf Integration of $({\cal L}_\nu)^\rho$:}\\
The integral
\begin{eqnarray}
({\cal L}_\nu)^{\rho} &=& \int 
\frac{d^3 p_f}{E_f}
\frac{d^3 p_\nu}{E_\nu} 
~\delta^{(4)}(Q - P_f - P_{\nu}) (P_\nu)^\rho 
\nonumber\\
\end{eqnarray}
can be written as 
\begin{eqnarray}
({\cal L}_\nu)^\rho = A Q^\rho.
\end{eqnarray}
Here, $A\equiv A(Q^2)$ is a scalar function of $Q^2$.
By multiplying $({\cal L}_\nu)^\rho$ with $Q_\rho$ 
the constant $A(Q^2)$ can be determined. Then we
get
\begin{eqnarray}
({\cal L}_f)^\rho  =  \frac{(Q\cdot P_f )}{ Q^2}
 ~{\cal K}~~  Q^\rho.
\end{eqnarray}

\vspace{0.2cm}

{\bf Integration of $({\cal N}_{\nu f})^{\rho\sigma}$}:\\
The integral
\begin{eqnarray}
({\cal N}_{\nu f})^{\rho\sigma} &=& \int 
\frac{d^3 p_f}{E_f}
\frac{d^3 p_\nu}{E_\nu} 
~\delta^{(4)}(Q - P_f - P_{\nu}) (P_\nu)^\rho (P_f)^\sigma
\nonumber\\
\end{eqnarray}
is a second rank tensor
\begin{eqnarray}
({\cal N}_{\nu f})^{\rho\sigma} 
= C g^{\rho\sigma} + D  Q^{\rho} Q^{\sigma},
\end{eqnarray}
where $C\equiv C(Q^2)$ and  $D\equiv D(Q^2)$ are 
scalar functions of $Q^2$. 

By multiplying $({\cal N}_{\nu f})^{\rho\sigma}$ 
with $g^{\mu\nu}$ and with  $Q_\rho Q_\sigma$
a set of two equations is formed. By solving them we find 
\begin{eqnarray}
\frac{({\cal N}_{\nu f})^{\rho\sigma}}{\cal K} &=&
\left(
\left( P_\nu\cdot P_f \right) - 
\frac{\left( Q\cdot P_\nu \right) \left( Q\cdot P_f \right)}
{Q^2} \right) 
~\frac{1}{3}  g^{\rho\sigma} 
\nonumber\\
 && -  \left(
\left( P_\nu\cdot P_f \right) - 
4 \frac{\left( Q\cdot P_\nu \right) \left( Q\cdot P_f \right)}
{Q^2} \right) 
\frac{Q^\rho Q^\sigma}{3 Q^2}.
\nonumber\\  
\end{eqnarray}

The remaining  integrals $({\cal L}_\nu)^\rho$, 
$({\cal N}_{\nu\nu})^{\rho\sigma}$, $({\cal N}_{f f})^{\rho\sigma}$
can be calculated following the scheme given above.

\end{document}